\documentclass{elsart}
\usepackage{epsfig}
\usepackage{graphicx}
\usepackage{dcolumn}
\usepackage{bm}
\usepackage{amsmath}
\usepackage{amsfonts}
\usepackage{amssymb}
\begin{document}
\begin{frontmatter}
\title{Need, Greed and Noise: Competing Strategies in a Trading Model}
\author{R. Donangelo,\thanksref{UFRJ}}
\author{A. Hansen,\thanksref{tron}}
\author{K. Sneppen, and }
\author{S. R. Souza\thanksref{UFRJ}}
\address{The Niels Bohr Institute,\\
Blegdamsvej 17, Copenhagen \O, Denmark}
\thanks[UFRJ]{Permanent address: Instituto de F\'{\i}sica,
Universidade Federal do Rio de Janeiro,
C.P. 68528, 21941-972 Rio de Janeiro, Brazil}
\thanks[tron]{Permanent address: Department of Physics,
Norwegian University of Science and Technology,
N-7491 Trondheim, Norway}
\date{\today}
\begin{abstract}
We study an economic model where agents trade a variety of products
by using one of three competing rules: ``need", ``greed" and
``noise". We find that the optimal strategy for any agent depends
on both product composition in the overall market and composition
of strategies in the market. In particular, a strategy that does
best on pairwise competition may easily do much worse when all
are present, leading, in some cases, to a ``paper, stone, scissors" 
circular hierarchy.
\end{abstract}
\end{frontmatter}

\section {\bf Introduction}

Human activity often takes the form of exchanges. These exchanges typically
consist of goods that can be quantified by value, but also opinions or
other types of information may be traded. The former define a market economy.
There have been several proposals to model such markets, see, for example the 
review by Farmer \cite{farmer}.
Most of the proposed models aim at reproducing the fat tails and volatility 
clustering in a stock or currency market
\cite{lux,bornholdt,bouchaud,stauffer,paczuski}.
Earlier we have proposed a market model (``Fat Cat" model) where agents 
trade products according to individual price estimates.  These estimates were 
dynamically adjusted as a function of the trading encounters of each agent \cite{DHSS}. 

The model mentioned above is one of many similar models that could be considered 
for such a market, each model being distinguished by a strategy that defines which 
product a buyer should select from a seller and how he should price it later.
In the present paper we study the interplay of a few strategies. The relative 
performance of these strategies is quantified by the wealth of agents employing 
them.

We organized the paper by first reviewing the ``Fat Cat" model in section II. 
The extension of this model to the case of other strategies is described in 
section III. Then, in section IV we discuss how the model could be extended to 
cases where agents are able to change their strategies according to their 
performance and give our concluding remarks.

\section {\bf Rules of the game}

We picture our model as a cartoon of the trading situations found in a real market. 
As shown in fig. 1, this minimalistic model consists of a system where agents trade 
a set of $N_{pr}$ different products. Each agent $i$ is assigned one of three possible 
strategies selected to be either based on profit (i.e.\ ``greed", which is the
strategy adopted in the original ``Fat Cat" model), on the need for a particular
product, or on a random selection without regard to the level of profit or need.

\begin{figure}
\centerline{\psfig{file=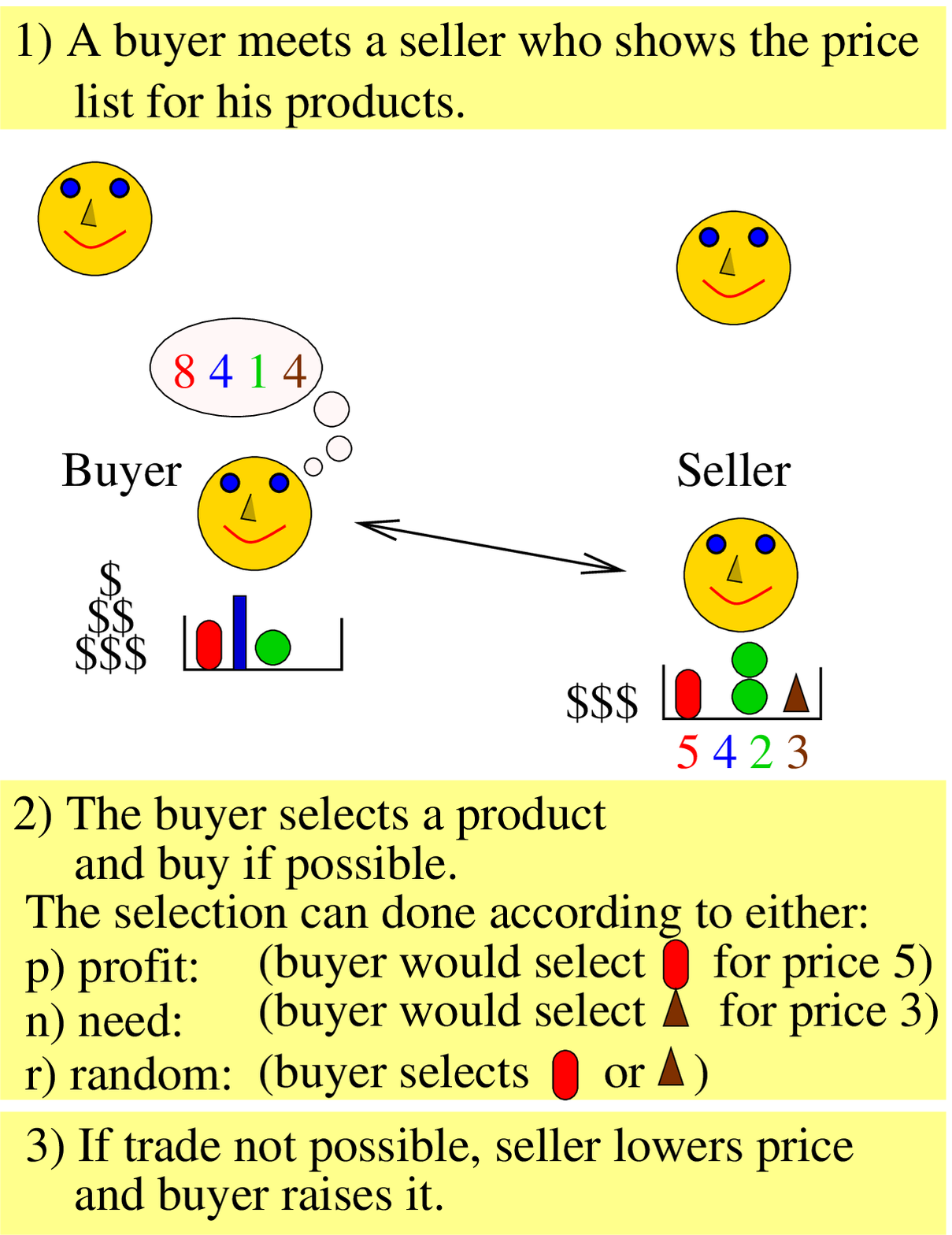,angle=0,width=8cm,clip=}}
\vspace{0.5cm} \caption[] {\small\sl Pictorial representation of the model. A buyer 
and a seller in a market with $N_{pr}=4$ products meet.  The buyer has one of products 
1 and 3, two of product 2 and none of product 4.  The seller, has one of products 1 and 4, 
two of product 3 and none of product 2.  The buyer values the four products at 8, 4, 1 and
4 units of money and the seller values them at 5, 4, 2 and 3, respectively.
The buyer compares his price list for the four products with the seller's price list, and 
according to strategy proposes a deal.} \label{money}
\end{figure}
\vspace{0.5cm}

Other strategies could be explored; for example, agents could act as ``garbage collectors", 
buying whatever product has the lowest possible price. In this paper we limit ourselves to 
the three, probably most basic, strategies outlined in fig. 1.

Each of the $N_{ag}$ agents starts with $N_{un}$ units of goods.  The goods are randomly
selected, for each agent, among the $N_{pr}$ different products.  Thereby we form the stock 
$S$ of each agent that together with some initial amount of money, $N_{mon}$ define the 
initial state of the economy.  We describe below the dynamics that arises from the 
interactions among these agents.

During the time evolution of the system, each agent $i$ has, at each time step, an amount of 
money $M(i), i=1,...,N_{ag}$, and a stock of the different products $j,\, S(i,j)$, where
$j=1,...,N_{pr}$. The prices of the different items in the stock of agent $i$ are denoted 
$P(i,j)$ which initially are taken to be integers uniformly drawn in the interval $[1,5]$.
In all cases we have verified that the evolution of the system does not depend on this 
particular choice.
Agents then meet and exchange products and adjust prices. Price adjustment is such
that large differences in pricing between agents are lowered, but also such that price 
differences are induced by some noise when they are small, as in real markets.

As in our simulations for the original ``Fat Cat" model, we assume that at each 
time step the following procedure takes place:

\begin{itemize}
\item{\bf 1} Buyer $(b)$ and seller $(s)$ are selected at random among
the $N_{ag}$ agents. If the seller has no products to offer, then
another seller is chosen.

\item{\bf 2} The buyer selects a product $j$ in the seller's stock
according to his strategy $\sigma(i)$.
\par \noindent i) If strategy $\sigma(i)$ is ``Profit" then he selects the product
$j$ which maximizes $P(b,j)-P(s,j)$, (i.e. his profit).
\par \noindent ii) If strategy $\sigma(i)$ is ``Need" then he selects the product
$j$ which he has the least in his stock (minimizes $S(i,j)$).
\par \noindent iii) If strategy $\sigma(i)$ is ``Random" then he selects a random
product $j$ in the seller's stock for which $P(b,j)-P(s,j)>0$.
\par \noindent The selected product $j$ is called $j_{bb}$ (best buy).

\item{\bf 3a}  If the buyer does not have enough money, (i.e. if $M(b)<P(b,j_{bb})$, 
we return to the first step and choose a new pair of agents.

\item{\bf 3b} If the buyer has enough money we proceed.
\par \noindent If $P(s,j_{bb})<P(b,j_{bb})$, the transaction is performed
at the seller's price. This means that we adjust:
$S(b,j_{bb})\rightarrow S(b,j_{bb})+1$,
$S(s,j_{bb})\rightarrow S(s,j_{bb})-1$,
$M(b)\rightarrow M(b)-P(s,j_{bb})$, $M(s)\rightarrow M(s)+P(s,j_{bb})$.

\item{\bf 3c} If $P(s,j_{bb}) \ge P(b,j_{bb})$, the transaction is not
performed. In this case, the seller lowers his price by one unit,
$P(s,j_{bb})\rightarrow \max (P(s,j_{bb})-1,0)$, and the buyer
raises his price by one unit,
\par \noindent $P(b,j_{bb})\rightarrow P(b,j_{bb})+1$.
\end{itemize}

The prices are always non-negative integers. Also note that since,
as defined in step 3 above, the price offered by the buyer cannot be
higher than the amount of money it has, we are not allowing for the
agents to get into debt. In case there are several products that
fulfill the selection criterion in 2, a random one of these is chosen.
One should emphasize that due to the price adjustments performed in
unsuccessful encounters, the prices never reach equilibrium, and
different agents typically assign different prices to the same product.

To quantify the system, we define the total wealth of an agent $i$ as the
amount of money plus the value of all goods in the agent's possession:
\begin{equation}
W(i)=M(i)+G(i)
\end{equation}
The value of product $j$ is defined as the average of what all agents
consider its value to be:
\begin{equation}
P_{ave}(j)=\frac{1}{N_{ag}}\sum_{i=1}^{N_{ag}} P(i,j),
\end{equation}
and the value of all of agent $i$'s goods, $G(i)$ is defined as
\begin{equation}
G(i)=\sum_j S(i,j) P_{ave}(j)\,.
\end{equation}
The ``Fat Cat" model's ``profit" rule was studied extensively
in our earlier paper. There it was found to lead to persistency and
fat tails in the agents' wealth fluctuations with time, hence its name.
We now study the other strategies and their interactions.

\section {\bf Fixed Strategies}

The three strategies given above lead to different wealth of the
respective agents. We will now study this in some detail. Our first
step is to illustrate how the wealth of the agents gets distributed
in a system where all agents employ the same strategy. For that
purpose we consider a system composed of 50 agents, 50 different
products. Each agent is given 40 units of products and 20 units of money. 
The result, shown in fig. 2, is that there are appreciable differences 
in the wealth distribution according to the strategy. For a profit-minded
system (the original ``Fat Cat" model), there is a long tail of
wealthy agents, which becomes less pronounced when the strategy is
based on the stock needs. In this case there are no rich agents,
but instead a large concentration of middle-wealth agents.
If the strategy adopted is to buy a random product, the long tail 
disappears and the middle-wealth peak is shifted to lower values than 
in the previous case.

\begin{figure}
\centerline{\psfig{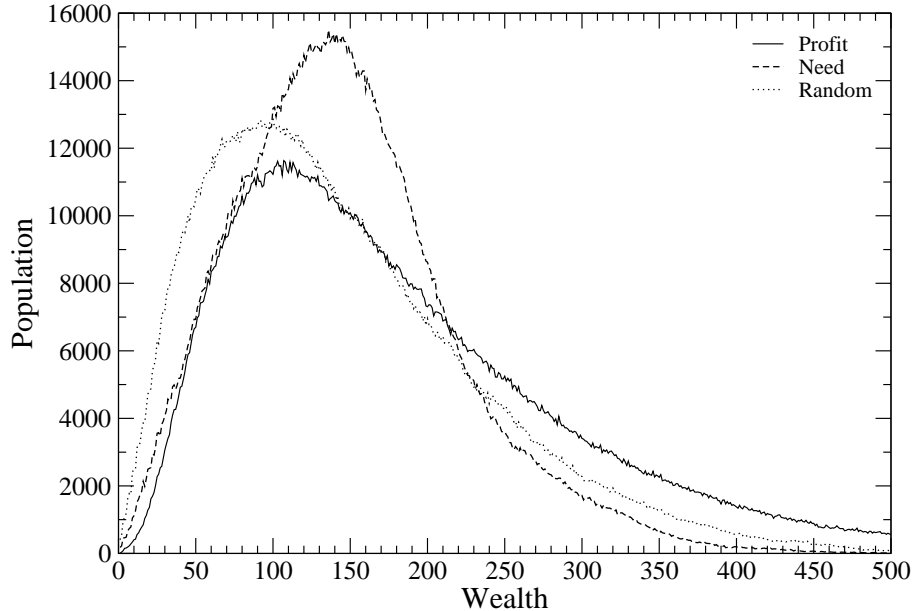}}
\vspace{0.5cm} \caption[] {\small\sl Wealth distribution for all agents
employing either of the three strategies described in the text. The
parameter values for the simulations were $N_{ag}=N_{pr}50$, and,
initially, $N_{un}=40$, $M=20$ per agent.
\vspace{1.0cm}}
\label{wealth}
\end{figure}

\begin{figure}
\centerline{\psfig{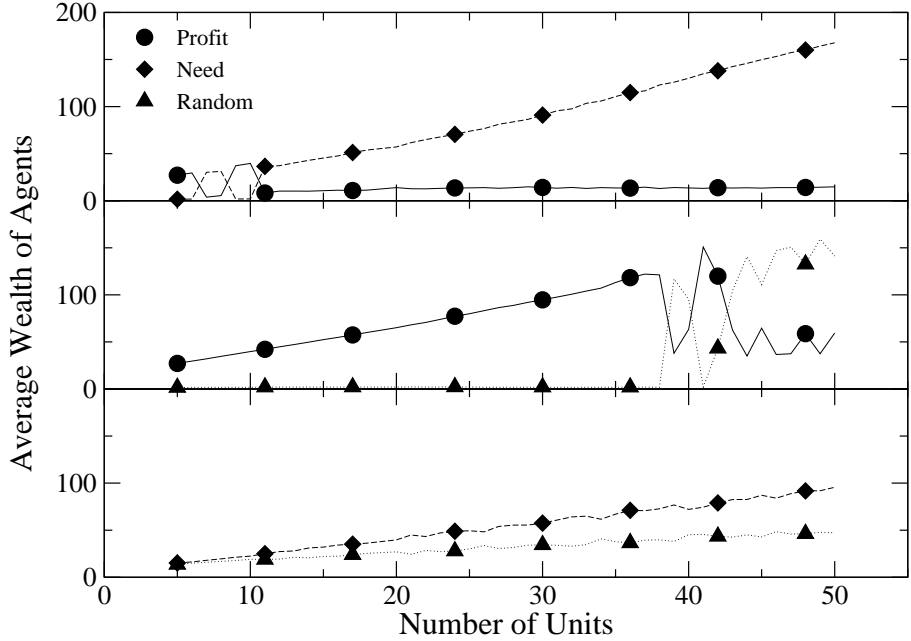}}
\caption[] {\small\sl Pairwise competition between strategies as a 
function of the number of units of goods in the limit when there are 
few copies of each good (``antique market"). In all cases the total
number of agents, products and initial units of money per agent are 
the same as in Fig. 2.}
\label{twostrat}
\end{figure}

The situation can change when the economic conditions, as defined
by the number of units of products and money in the economy, are
changing. We will show that there is no best rule for all situations
This is shown in fig. 3, where we have changed the number of units
of products, while keeping all other model values the same, in
markets composed of equal numbers of agents employing two different
strategies. In this case, where $N_{un} < N_{pr}$, each agent has a few
of many possible products, thus representing an antique dealer market
where there are many special items and no one can have everything.
We see, in the upper panel, that for this market the ``need" strategy,
is better than ``profit" above 10 units. However, below 10 units,
the result is unstable, one strategy leading to better results than
the other depending on the initial conditions. In the middle panel we 
see that the ``random" agents have a higher average wealth than the 
``profit" motivated ones when the number of units is above 45, the 
situation is reversed for a number of units below 35, and there is
unpredictability in between these two values. Finally, the lower 
panel shows that ``need" works better than a ``random" strategy
for all number of units in the range shown.
For other parameter values, namely when the number of units of
products and money is large, the ``random" strategy outperforms ``need".

In this last case, namely for $N_{un} >> N_{pr}$, each agent typically
has many copies of all products, which represent a mass production
market with many copies of few items; A supermarket world. In this
case, a pairwise comparison of the three strategies shows an interesting
situation, depicted in fig. 4: ``need" outperforms ``profit", which
outperforms ``random", which outperforms ``need". So, in this case
no strategy is better than both of the other strategies when studied pairwise.  

\begin{figure}
\centerline{\psfig{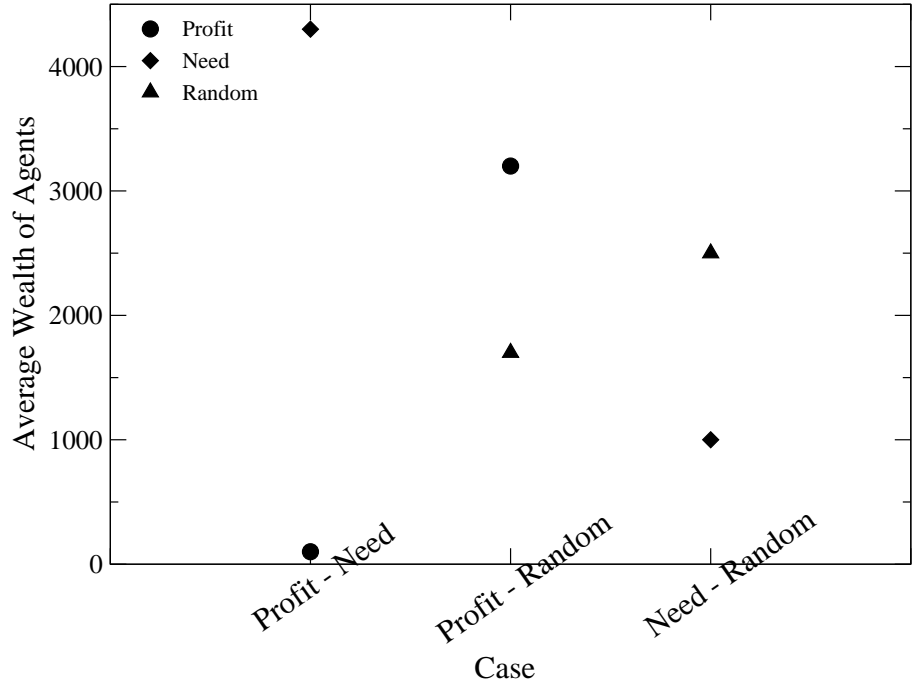}}
\vspace{0.5cm} \caption[] {\small\sl Pairwise competition between
strategies for a rich market, in which each agent has initially 
$N_{un}=M=400$.  The remaining parameters are the same as in Fig. 3.
Here we find the circular hierarchy ``need" $>$ ``profit", 
``profit" $>$ ``random", and ``random" $>$ ``need".}
\label{threestrat}
\end{figure}

In fig. 5 we explore the triple market further, implementing a market
where there is only one agent with ``need" strategy  among 25 agents 
with ``profit" and 25 agents with ``noise" as strategies, so that 
there are all together 51 agents.  Without the single ``need" strategy, 
``random" would  outperform ``profit" with a large margin. A single
``need" agent in the system will slowly collect a large fraction
of all products in the market. This is because it has a large competition
advantage over the ``profit", which overcompensates its disadvantage
to the ``random" ones. If the number of ``profit" agents were reduced
and that of the ``random" agents increased, the ``need" agent would
do worse. This illustrates the fact that the number of agents employing
the different strategies is also determinant in the relative success
of the agents adopting them.

\begin{figure}
\centerline{\psfig{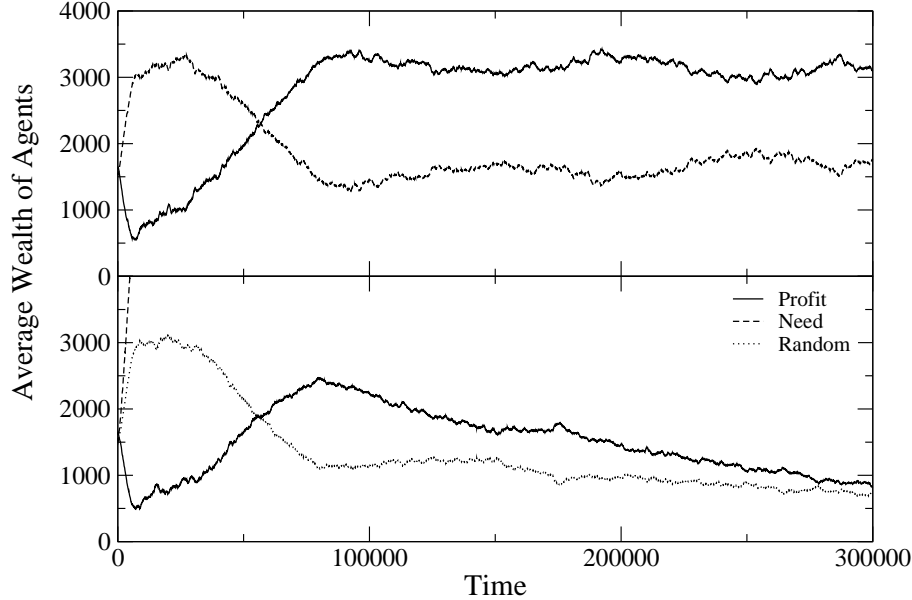}}
\vspace{0.5cm} \caption[] {\small\sl a) Competition between 25
agents employing the ``profit" strategy with a similar number
of ``random" agents, in the same rich market as in Fig. 4. 
b) Competition where a single ``need" agent is introduced into the 
system of the panel above. Note that this agent gathers so much 
wealth that he rapidly grows out of the scale of the illustration.}
\label{threefixstrat}
\end{figure}

\section {\bf Outlook and conclusions}

As we have seen, one may have different strategies for different
agents. An interesting direction to extend this model is to let
the agents to select the strategies they adopt in order to 
improve their performance. A further extension development
would be to allow the strategies themselves to evolve. In the
following we show a preliminary example of how the agents
could choose strategies.

A simple mechanism consists in updating, at fixed time intervals, 
the strategy of the poorest agent in the system. This agent just 
changes his present strategy to any of the other two, at random.
The resulting time evolution for a system composed of 45 agents initially
equally distributed according to their strategies is shown in fig. 6.
We see that, after an initial transient, the number of agents following
the ``need" and ``random" strategies becomes approximately equal and
constant for a considerable period of time, while the ``profit" strategy
is followed by a small fraction of the agents. An increase in the ``profit"
agents leads to a change in the conditions which lead the ``random"
strategy to almost disappear from the system, to a dominance of ``need",
and the persistency of ``profit" at a very low level. From the fig. 4
we see that were it not for the existence of ``profit"-thinking agents,
``random" would dominate over ``need".

The results shown in fig. 6 also demonstrate that the hierarchy
``paper --- stone --- scissors" illustrated in fig. 4, for the same
system, does not seem to lead, in the time interval considered, 
to alternations in the number of agents employing each strategy.
There could be several reasons for this. One is that these
changes take longer and longer time to alter enough the market
conditions to significantly modify the performance of the
different strategies. Another is that while in fig. 4 one measures the 
average wealth of the agents using each of the three strategies, the 
criterium for changing strategies is based on the performance of poor 
agents, {\it i.e.} on the agents' wealth distribution in the region 
close to the origin. As suggested by fig. 2, a strategy yielding a 
higher average wealth than another, may also have a larger number of 
poor agents than the other. These issues need to be further explored 
for an appropriate strategy selection procedure and will be discussed 
in the future.

\begin{figure}
\centerline{\psfig{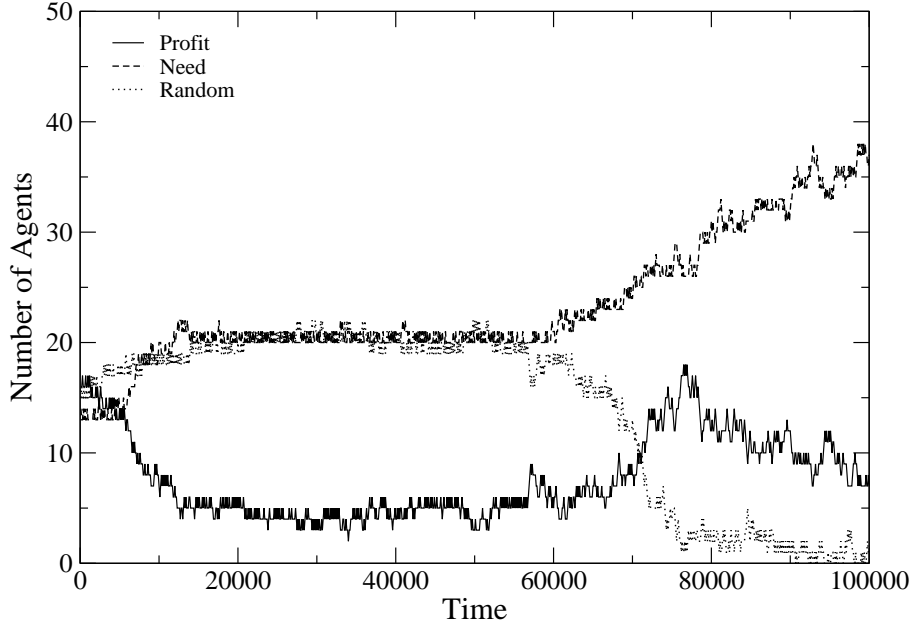}}
\vspace{0.5cm} \caption[] {\small\sl Number of agents employing
each or the three strategies as a function of time. The model
parameters are the same as in Figs. 4 and 5.}
\label{threevarstrat}
\end{figure}

Compared to earlier models of market dynamics the model presented here 
has some new and related key features: there is local optimization of 
utility (estimated market value) and all trades are done locally without 
the equilibrizing effects of a global information pool. 
This gives arbitrage possibility which drives a dynamic and evolving market. 
Earlier market models, such as the minority game, have a global information 
pool, and lack dynamical signals that could be associated with stock market 
fluctuations. The evolving Boolean network for minority games of Paczuski, 
Bassler and Corral \cite{paczuski}, on the other hand,  works with local 
information exchange, but the reward function is still global.
Also the frame of minority games makes it difficult to treat a multi-product 
market, which we believe is important for understanding real stock markets.

The model we propose is for a market composed of agents, goods and money
(or People, Prices and Products).
We have demonstrated that such a market easily shows persistent fluctuations 
of wealth with time, and seen that the persistency is closely related to an 
interplay of having many products that influence each others trade probability.
A similar result was obtained in the simpler model in \cite{DS99}, where it
was demonstrated that persistency could arise even without money.
The setup proposed here with agents and products with individual local prices
allows for individual strategies of the agents. This opens for evolution of
strategy as a part of the financial market, and we have seen that evolving
strategies indeed give a dynamics where wealth is often rapidly redistributed.

\vspace{0.5cm}

R.D. and S.R.S. acknowledge support from the Brazilian National 
Research Council (CNPq) and from the Niels Bohr Institute.


\vspace{0.5cm}

\begin{thebibliography}{99}
\bibitem{farmer} J.D. Farmer, \textit{Comp. in Science and Eng.}
\textbf{1} (6) 26 (1999).

\bibitem{lux} T. Lux and M. Marchesi, Nature {\bf 397} 498 (1999).

\bibitem{bornholdt} T. Kaizoji, S. Bornholdt and Y. Fujiwara, Physica A
{\bf 316} 441 (2002).

\bibitem{bouchaud} J.-P. Bouchaud and M. Potters, Theory of Financial Risk 
and Derivative Pricing, 2nd Edition, Cambridge University Press, 2003.

\bibitem{stauffer} I. Chang, D. Stauffer and R. B. Pandey,
J. Theor. Appl. Finance {\bf 5} 585 (2002).

\bibitem{paczuski} M. Paczuski, K. Bassler and A. Corral, Phys. Rev. Lett.
{\bf 84} 3185 (2000).

\bibitem{DHSS} R. Donangelo, A. Hansen, K. Sneppen and S.R. Souza,
\textit{Physica A} \textbf{283} 469 (2000).

\bibitem{DS99} R. Donangelo and K. Sneppen, 
\textit{Physica A} \textbf{276} 572 (2000).


\end{thebibliography}
\end{document}